\documentclass[journal]{IEEEtran}

\usepackage{etoolbox}
\makeatletter
\@ifundefined{color@begingroup}%
{\let\color@begingroup\relax
\let\color@endgroup\relax}{}%
\def\fix@ieeecolor@hbox#1{%
\hbox{\color@begingroup#1\color@endgroup}}
\patchcmd\@makecaption{\hbox}{\fix@ieeecolor@hbox}{}{\FAILED}
\patchcmd\@makecaption{\hbox}{\fix@ieeecolor@hbox}{}{\FAILED}

\usepackage{cite}
\usepackage{amsmath,amssymb,amsfonts}
\usepackage{algorithmic}
\usepackage{placeins}

\usepackage{graphicx}
\usepackage{textcomp}
\usepackage{algorithm}
\def\BibTeX{{\rm B\kern-.05em{\sc i\kern-.025em b}\kern-.08em
    T\kern-.1667em\lower.7ex\hbox{E}\kern-.125emX}}
\begin{document}
\title{Multi-frequency Electrical Impedance Tomography Reconstruction with Multi-Branch Attention Image Prior}
% \author{A, \IEEEmembership{Student Member, IEEE}, and B, \IEEEmembership{Member, IEEE}
\author{Hao Fang, \IEEEmembership{Student Member, IEEE}, Zhe Liu, \IEEEmembership{Graduate Student Member, IEEE}, Yi Feng, Zhen Qiu, \IEEEmembership{Member, IEEE}, Pierre Bagnaninchi, and Yunjie Yang, \IEEEmembership{Senior Member, IEEE}
\thanks{Hao Fang, Zhe Liu and Yunjie Yang are with the SMART Group, Institute for Imaging, Data and Communications, School of Engineering, The University of Edinburgh, Edinburgh, UK. (Correspondence authors: Yunjie Yang and Zhe Liu; Email: y.yang@ed.ac.uk and zz.liu@ed.ac.uk).}
\thanks{Yi Feng is with the Centre for Inflammation Research, Queen's Medical Research Institute, The University of Edinburgh, Edinburgh, UK.}
\thanks{Zhen Qiu is with the School of Science and Engineering, University of Dundee, Dundee, UK.}
\thanks{Pierre Bagnaninchi is with the Centre for Regenerative Medicine, Institute for Regeneration and Repair, The University of Edinburgh, UK.}
}

\maketitle

\begin{abstract}
Multi-frequency Electrical Impedance Tomography (mfEIT) is a promising biomedical imaging technique that estimates tissue conductivities across different frequencies. Current state-of-the-art (SOTA) algorithms, which rely on supervised learning and Multiple Measurement Vectors (MMV), require extensive training data, making them time-consuming, costly, and less practical for widespread applications. Moreover, the dependency on training data in supervised MMV methods can introduce erroneous conductivity contrasts across frequencies, posing significant concerns in biomedical applications.
To address these challenges, we propose a novel unsupervised learning approach based on Multi-Branch Attention Image Prior (MAIP) for mfEIT reconstruction. 
Our method employs a carefully designed Multi-Branch Attention Network (MBA-Net) to represent multiple frequency-dependent conductivity images and simultaneously reconstructs mfEIT images by iteratively updating its parameters. By leveraging the implicit regularization capability of the MBA-Net, our algorithm can capture significant inter- and intra-frequency correlations, enabling robust mfEIT reconstruction without the need for training data. Through simulation and real-world experiments, our approach demonstrates performance comparable to, or better than, SOTA algorithms while exhibiting superior generalization capability. These results suggest that the MAIP-based method can be used to improve the reliability and applicability of mfEIT in various settings.
\end{abstract}


\section{Introduction}
\label{sec:introduction}
\IEEEPARstart{B}{ioimpedance} refers to the electrical impedance of biological tissues measured as current passes through them. It varies with frequencies, tissue types, and physiological status, and is sensitive to variations in underlying biology. \cite{b1,b2,b3}. 
% Electrical Impedance Tomography (EIT) reconstructs conductivity images by injecting currents through boundary electrodes and measuring induced voltages \cite{b38}. Its low cost, portability, non-invasiveness, radiation-free nature, and high temporal resolution make it increasingly popular in tissue and cellular imaging. 
 Multi-frequency EIT (mfEIT) reconstructs multiple frequency-dependant conductivity images from a series of voltage measurements rapidly, non-intrusively, and without radiation. Compared to single-frequency EIT (sfEIT), which only reconstructs the conductivity at a specific frequency, mfEIT provides more comprehensive insights into the physiological or pathological status of tissues. mfEIT applications include the detection of intracranial abnormalities\cite{b14,b15}, analysis of lung pathologies\cite{b13}, and monitoring of cell culture\cite{b45}. Despite its significant potential in medical imaging, the practical application of mfEIT is currently largely limited by low image quality.

%Most efforts to improve mfEIT image quality have focused on enhancing reconstruction algorithms. Based on how measurement data is utilized, 
Existing mfEIT reconstruction algorithms can be categorized into Single Measurement Vector (SMV)-based and Multiple Measurement Vector (MMV)-based methods.
% The intuitive method for mfEIT reconstruction is to use Single Measurement Vector (SMV)-based frameworks\cite{b41,b16,b17,b18}, 
SMV-based approaches treat mfEIT reconstruction as a series of single-frequency tasks, reconstructing the image at each frequency separately. This allows the use of various single-frame image reconstruction algorithms, including model-based iterative algorithms like Structure-Aware Sparse Bayesian Learning (SA-SBL)\cite{b16}, model-based learning methods such as ISTA-Net\cite{b21}, FISTA-Net\cite{b20}, and MoDL\cite{b19}, and unsupervised learning approaches like DeepEIT\cite{b18}. However, SMV-based methods do not account for inter-frequency correlations among mfEIT images, usually leading to structural inconsistencies across frequencies, inaccurate conductivity prediction, and increased noise vulnerability.

Multiple Measurement Vectors (MMV)-based methods\cite{b7,b9,b11}, on the other hand, reconstruct multiple conductivity images simultaneously by optimizing a multi-task objective function built using measurements from all frequencies. By exploiting shared features embedded across different frequencies, MMV-based methods effectively improve inter-frequency correlations. Notable model-based algorithms include
ADMM-MMV \cite{b9} and MMV-SBL \cite{b11}. While these methods have made progress in capturing inter-frequency correlations, they still yield unsatisfactory reconstruction quality and require extensive manual tuning of multiple parameters, thereby limiting their practical applicability.

% Many model-based algorithms developed within the MMV framework have shown promising performance 
% in various tomographic imaging tasks. Notable examples include Qu et al.\cite{b7}, Zhang et al.\cite{b8}, and Chen et al.\cite{b9}, who employed the ADMM-MMV approach in hyperspectral imaging, multi-frequency Electrical Capacitance Tomography (mfECT), and mfEIT, respectively. Similarly, Liu et al.\cite{b11} and Xiang et al.\cite{b12} respectively applied MMV-SBL for mfEIT and multi-frequency Electromagnetic Tomography (mfEMT). 
%Overall, these methods formulate reconstruction as a joint sparse recovery problem, minimizing a cost function that balances data fidelity and sparsity across multiple measurements, solved using iterative techniques like ADMM or Bayesian inference. 

Recently, supervised learning-based methods\cite{b40} have demonstrated superior performance in solving inverse problems within the MMV framework. These approaches can learn robust priors from large-scale datasets due to the excellent feature representation and non-linear fitting capabilities of carefully designed neural networks \cite{b26}. Representative methods include end-to-end learning algorithms like SFCF-Net\cite{b35} and model-based supervised learning algorithms such as MMV-Net\cite{b9}. By integrating neural networks into iterative steps, model-based supervised learning approaches combine the network's nonlinear fitting with the physical insights of model-based algorithms, offering superior generalization than end-to-end methods.
% However, these approaches heavily depend on the quantity and quality of training data, which is often difficult and expensive to obtain, particularly in the biomedical field.
However, the data reliance of model-based supervised learning algorithms still degrades their generalization ability. For instance, as demonstrated in Section V, MMV-Net reconstructs incorrect conductivities across different frequencies. 
% Consequently, these limitations hinder the broader application of learning-based mfEIT imaging.

In the most recent advances, dataset-free unsupervised learning methods\cite{b23}, exemplified by Deep Image Prior (DIP)\cite{b22}, have shown promising performance in inverse problems. These methods leverage the neural network's inherent structure and inductive biases to regularize the inverse problem, capturing essential image statistics without explicit priors. For instance, Gong et al.\cite{b24} and Ote et al.\cite{b25} applied DIP-based approaches to Positron Emission Tomography (PET) reconstruction, demonstrating improved image quality and noise resistance. In EIT, unsupervised learning methods have been explored for sfEIT reconstruction, outperforming traditional regularization-based approaches \cite{b18, b32, b34}. However, to the best of our knowledge, unsupervised learning methods have not yet been reported for mfEIT reconstruction. 

Here, we introduce the first unsupervised method for mfEIT reconstruction, aiming to capture robust inter- and intra-frequency correlations while improving image quality and generalization capablity. Our method introduces the neural network prior via representing multiple mfEIT images by a Multi-Branch Attention Network (MBA-Net). The MBA-Net features multiple branch subnetworks to capture multi-branch features from different frequency measurements, followed by a Fusion Unit (FU) and a Branch Attention (BA) modules to enhance the inter- and intra-frequency correlations. The mfEIT images are reconstructed by iteratively updating the MBA-Net parameters. We refer to this prior as the Multi-Branch Attention Image Prior (MAIP). 
Simulations and real-world experiments validate the proposed approach, demonstrating its superior performance among given algorithms. Our main contributions are as follows:

\begin{enumerate}
    \item We pioneer a model-based unsupervised learning method for mfEIT image reconstruction that excels in preserving imaging targets' structure, improving conductivity estimation accuracy, capturing inter-frequency correlations, and enhancing generalization capability.
    \item We propose the MBA-Net within the MAIP framework, featuring multiple branch subnetworks, a FU, and a BA module. The multi-branch structure, along with the carefully designed subnetworks, is tailored to enhance intra-frequency correlations, while the FU and BA modules effectively capture inter-frequency correlations.
    \item The proposed MAIP-based approach introduces a robust implicit regularization strategy, enabling mfEIT to adapt to various scenarios without relying on manually designed explicit priors, thereby expanding its potential applications.
\end{enumerate}

%The remainder of this paper is organized as follows: Section II introduces the principles of mfEIT. Section III details our proposed MAIP algorithm. Section IV outlines the experimental setup. Section V presents and analyzes the results from both simulations and real-world experiments. Finally, Section VI concludes the paper and discusses future work.

\section{mfEIT Image Reconstruction}
The objective of mfEIT image reconstruction is to reconstruct multiple conductivity images from a series of measurements taken at different frequencies.
% , revealing the frequency-dependent bioimpedance properties of the tissues.beginning with the formation of the mfEIT forward model. 
This task begins with the formulation of the mfEIT forward model. We adopt its linearized version:
\begin{equation}
\mathbf{V} = \mathbf{J\pmb{\Sigma}},
\label{e1}
\end{equation}
where $\mathbf{V}=[\Delta\mathbf{v}_{f_{1}},\Delta\mathbf{v}_{f_{2}},\ldots,\Delta\mathbf{v}_{f_{i}},\ldots,\Delta\mathbf{v}_{f_{L}}]\in\mathbb{R}^{M\times L}$, and $\mathbf{\Sigma}=[\Delta\pmb{\sigma}_{f_{1}},\Delta\pmb{\sigma}_{f_{2}},\ldots,\Delta\pmb{\sigma}_{f_{i}},\ldots,\Delta\pmb{\sigma}_{f_{L}}]\in\mathbb{R}^{N\times L}$ are the normalized voltage measurement matrix and conductivity matrix, respectively. $\mathbf{J}\in\mathbb{R}^{M\times N}$ represents the normalized sensitivity matrix \cite{b33}. $i=1,2,...,L$ denotes the $i\text{-th}$ observation frequency, and $L$ stands for the total number of frequencies. $M$ is the number of measurements and $N$ is the number of pixels in a mfEIT image. 

\begin{figure}[!t]
	\centerline{\includegraphics[scale=0.11]{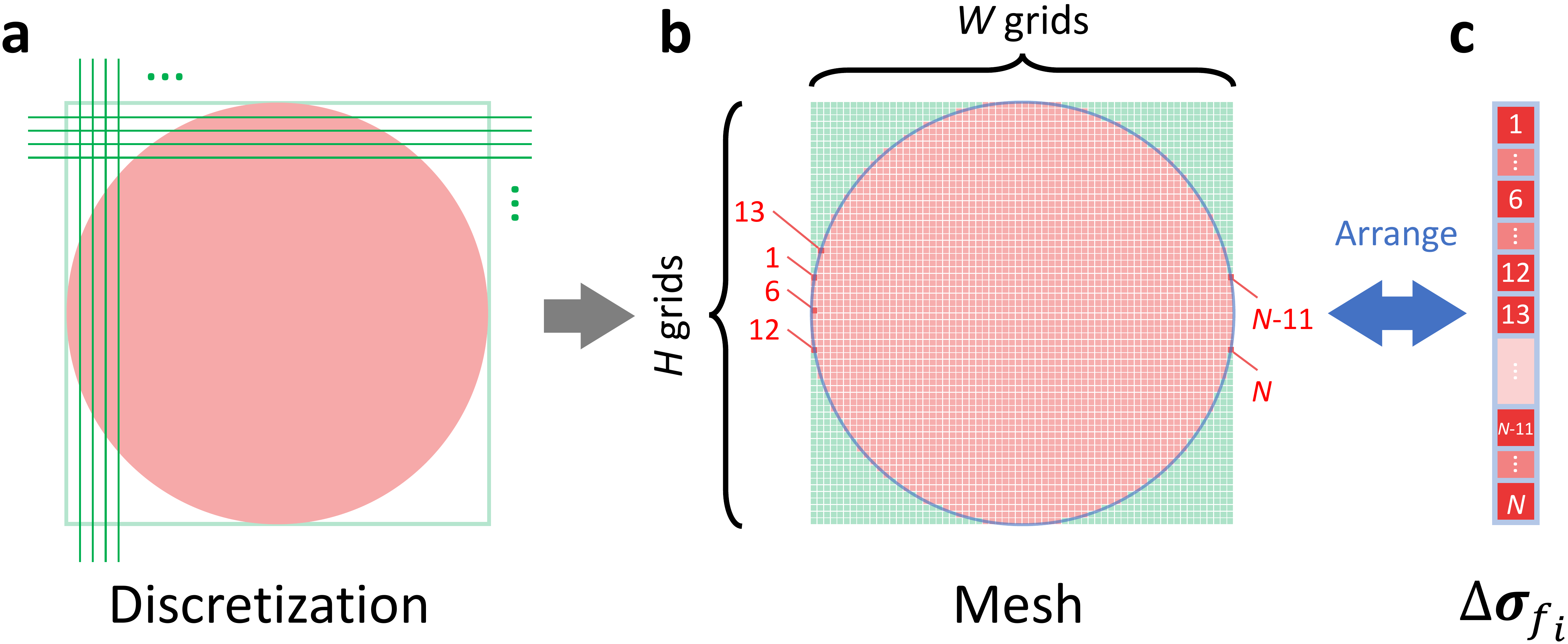}}
	\caption{Illustration of the imaging region discretization and the conversion between $\Delta\pmb{\sigma}_{f_{i}}$ and its corresponding rectangular image representation: The circular imaging region in ($\pmb{\textrm{a}}$)  is discretized into an \textit{H}-by-\textit{W} mesh in ($\pmb{\textrm{b}}$), where each pink grid in the mesh represents a pixel and the green grids denote the void pixels. The pink grids are indexed and can be further arranged into a vector $\Delta\pmb{\sigma}_{f_{i}}$ ($\pmb{\textrm{c}}$). Conversely, $\Delta\pmb{\sigma}_{f_{i}}$ can be rearranged into its rectangular form.}
    \label{mapping}
\end{figure}

There are two common imaging strategies in mfEIT: Time-Difference (TD) and Frequency-Difference (FD) imaging. The key distinction lies in the selection of the reference measurements. In TD-mfEIT,  reference measurements for a specific observation frequency are collected from the background medium using current stimulation at the same frequency. In contrast, FD-mfEIT uses measurements taken at a fixed frequency, with the imaging objects present, as the reference.

Based on (1), mfEIT reconstruction can be formulated as the following optimization problem: 

\begin{equation}
\min_{\mathbf{{\Sigma}}} \left\|\mathbf{{J}{\Sigma}}-\mathbf{V} \right\|+R(\mathbf{{\Sigma}}),
\end{equation}
where $\|\cdot\|$ represents a chosen norm to quantify data fidelity, with common examples being the Frobenius norm, $l_{1}$ norm, and so on. $R:~\mathbb{R}^{N \times L} \rightarrow \mathbb{R}$ denotes the regularization function, which embeds prior knowledge into the inversion. 

\section{Methodology}
In this section, we first propose a modified mfEIT forward model, adapted for tensor reshaping to ensure compatibility with the tensor framework. We then describe the mfEIT image reconstruction based on the Multi-Branch Attention Image Prior (MAIP). Finally, we provide details of the neural network architecture employed in the MAIP.

\subsection{Modified mfEIT Forward Model}
% Conventionally, images are processed and displayed in the form of two-dimensional (2D) matrices. However, 
We adopt the rectangular inverse mesh as described in \cite{b33}. In this case, arranging the $\Delta\pmb{\sigma}_{f_{i}}$ into a 2D rectangular image may result in some void or undefined pixels, as the shape of the mfEIT imaging region is typically non-rectangular (see Fig.~\ref{mapping}). Therefore, the 2D representation of $\Delta\pmb{\sigma}_{f_{i}}$ is non-structured and cannot be obtained using tensor reshaping. However, as we will demonstrate in the following subsections, the MAIP-based algorithm requires direct manipulation of the structured 2D version of $\Delta\pmb{\sigma}_{f_{i}}$ using tensor operations. To accommodate this need, we need to modify the original mfEIT forward model in (1) while preserving its mathematical interpretation, and enabling the tensor reshaping to the modified $\Delta\pmb{\sigma}_{f_{i}}$ and $\mathbf{\Sigma}$. Therefore, we formulate the following modified mfEIT forward model:
% To maintain mathematical consistency and enable the network to process the calculated conductivity distribution vector $\Delta\mathbf{\sigma}\in\mathbb{R}^{N}$ in a two-dimensional format, thereby enhance its ability to extract spatial relationships and local structural features, we modify the mfEIT forward model \eqref{e1} into \eqref{e2} using a mapping matrix $\mathbf{M}\in\mathbb{R}^{N\times (H \times W)}$. $\mathbf{M}$ is calculated based on the sensor setup, mapping the 3228 sensor pixels within a circular region to a 64x64 pixel 2D image grid (4096 pixels). Each sensor pixel's position is translated into the 2D grid using its row and column indices, setting the corresponding element in $\mathbf{M}$ to 1. 
\begin{equation}
\mathbf{V} = \mathbf{\widetilde{J}\widetilde{\Sigma}}, 
\label{e2}
\end{equation}
where $\mathbf{\widetilde{J}}=\mathbf{JP^{T}}\in\mathbb{R}^{P\times (H \times W)}$ and $\mathbf{\widetilde{\Sigma}}=\mathbf{P\Sigma}\in\mathbb{R}^{(H \times W)\times L}$. $\mathbf{P}\in\mathbb{R}^{(H \times W)\times N}$ is a custom projection tensor that satisfies $ (H\times W) \geq N$ with the condition $\mathbf{P^{T}P=I}\in\mathbb{R}^{N\times N}$, where $\mathbf{I}$ is an identity tensor. $H$ and $W$ denote the height and the width of the 2D rectangular image arranged from $\Delta\pmb{\sigma}_{f_{i}}$. From an intuitive standpoint, the projection tensor introduces zero values into the $\Delta\pmb{\sigma}_{f_{i}}$, thereby substituting all undefined or void elements (green grids in Fig.~\ref{mapping}b) in the original rectangular representation of $\Delta\pmb{\sigma}_{f_{i}}$ with zero values.

% To calculate the mapping tensor $\mathbf{M}$, a position tensor $\mathbf{P}\in\mathbb{R}^{N\times 2}$ is required. $\mathbf{P}$ consists of $N$ pairs of row and column indices, which are used to determine the pixel positions in an $H \times W$ 2D EIT image domain corresponding to the one-dimensional conductivity distribution.
% \begin{equation}
%     \mathbf{M}_{a,b}=\begin{cases}1,&\text{if}\quad a=(\text{col}_b-1)\times H+\text{row}_b,\\0,&\text{otherwise}.\end{cases}
% \end{equation}
% where $a=1,2,...,(H \times W)$ and $b=1,2,...,N$ represent the flattened pixel index in the $H\times W$ 2D image domain and the index of the conductivity data points, respectively. $\text{row}_b$ and $\text{col}_b$ are the row and column indices from the 
% $b\text{-th}$ row of the position tensor 
% $\mathbf{P}$.

Based on (3), the mfEIT reconstruction is formulated as:
\begin{equation}\arg\min_{\mathbf{\widetilde{\Sigma}}}\left\|\mathbf{\widetilde{J}\widetilde{\Sigma}}-\mathbf{V}\right\|+ \widetilde{R}\left(\mathbf{\widetilde{\Sigma}}\right),
\end{equation}
where $\widetilde{R}$ represents the mapping from $\mathbb{R}^{(H \times W) \times L}$ to $\mathbb{R}$.

\subsection{MAIP-based mfEIT Reconstruction}
% Specifically, we propose a Multi-Branch Attention Image Prior (MAIP) framework that leverages the regularization properties of our carefully designed Multi-Branch Attention Network (MBA-Net) for mfEIT image reconstruction. Fig.~\ref{MAIP} illustrates the flowchart of our MAIP framework for mfEIT reconstruction. This framework incorporates a Multi-Branch Network (MBA-Net) composed of multiple branch subnetworks, a Fusion Unit (FU), and a Branch Attention (BA) module (see Fig. 2).  

In MAIP, we represent the unknown multi-frequency conductivity distribution $\mathbf{\widetilde{\Sigma}}$ by a deep neural network, i.e.
\begin{equation}
\mathbf{\widetilde{\Sigma}}=\mathcal{R}_{v}\left(\phi\left(\pmb{\theta}|\mathbf{Z}\right)\right),
\end{equation}
where $\phi: \mathbb{R}^{L\times H\times W} \to \mathbb{R}^{L\times H\times W}$ stands for the proposed Multi-Branch Attention Network (MBA-Net). $\pmb{\theta}$ denotes parameters of the MBA-Net, and $\mathbf{Z}\in\mathbb{R}^{L\times H\times W}$ is the input noise tensor sampled from a uniform distribution $\mathbf{Z}\sim U(0,1)$. $\mathcal{R}_{v}:\mathbb{R}^{L\times H\times W}\to\mathbb{R}^{(H\times W)\times L}$ denotes the reshaping operation that reshapes a third-order tensor into a second-order tensor.

Substitute (5) to (4) and discard the regularization term $\widetilde{R}(\mathbf{\widetilde{\Sigma}})$, the MAIP-based reconstruction is formulated as the following nonlinear optimization problem:
\begin{equation}\label{prob}
\arg\min_{\pmb{\theta}}\left\|\mathbf{\widetilde{J}} \mathcal{R}_{v}(\phi(\pmb{\theta}|\mathbf{Z}))-\mathbf{V}\right\|_1,
\end{equation}
where,
\begin{equation}
\left\|\mathbf{\widetilde{J}}\mathcal{R}_{v}(\phi(\pmb{\theta}|\mathbf{Z})) - \mathbf{V}\right\|_1 = \sum_{i=1}^M \sum_{j=1}^L \left|\left(\mathbf{\widetilde{J}} \mathcal{R}_{v}(\phi(\pmb{\theta}|\mathbf{Z}))\right)_{ij} - V_{ij}\right|.
\end{equation}
%which represents the $\ell_1$ norm.

\begin{figure}[!t]\label{FC}
	\centerline{\includegraphics[scale=0.31]{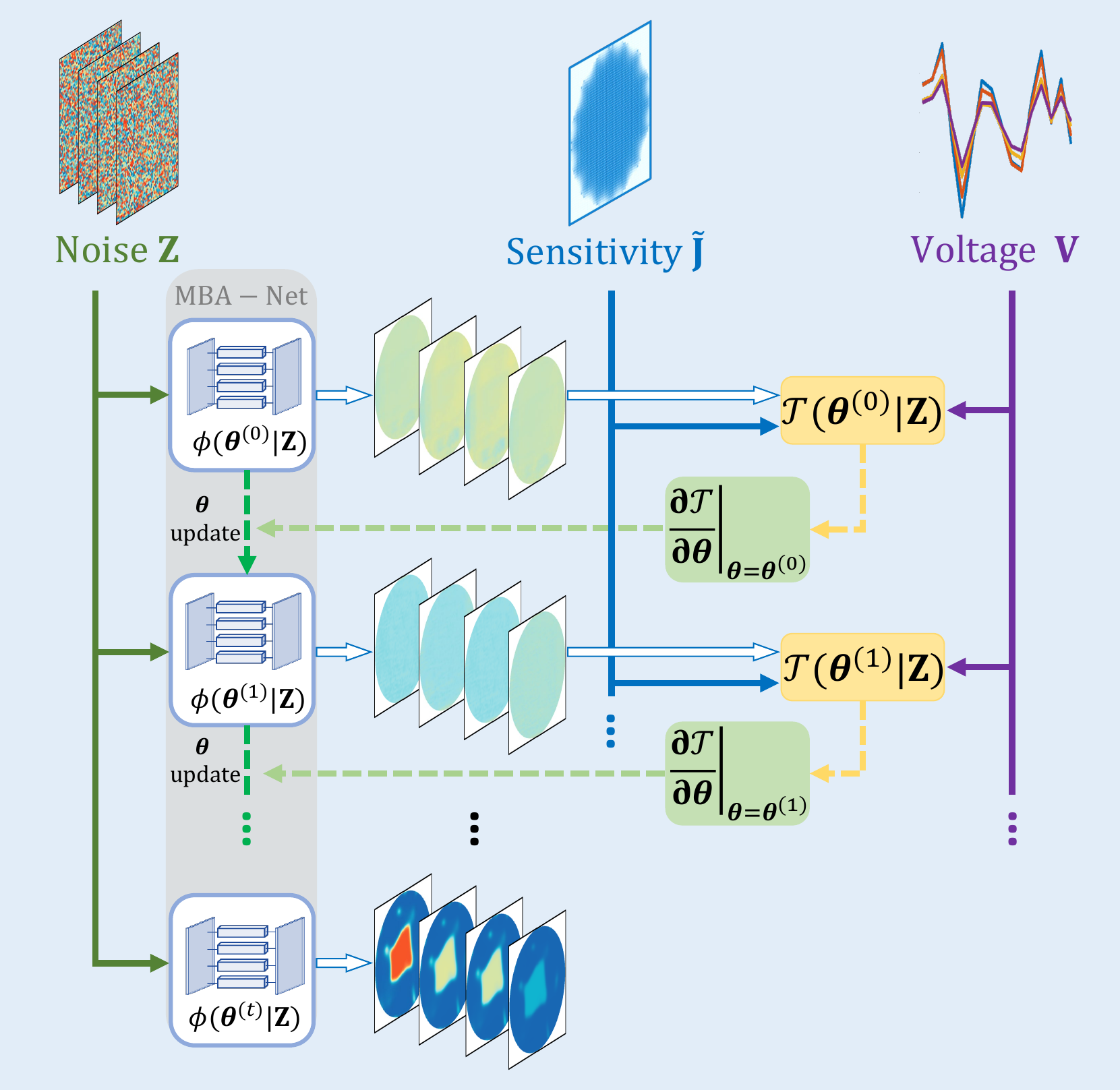}}
	\caption{Schematic of the MAIP-based mfEIT reconstruction approach.}
	\label{MAIP}
\end{figure}

\begin{figure*}[htbp]
\centering
\includegraphics[scale=0.4]{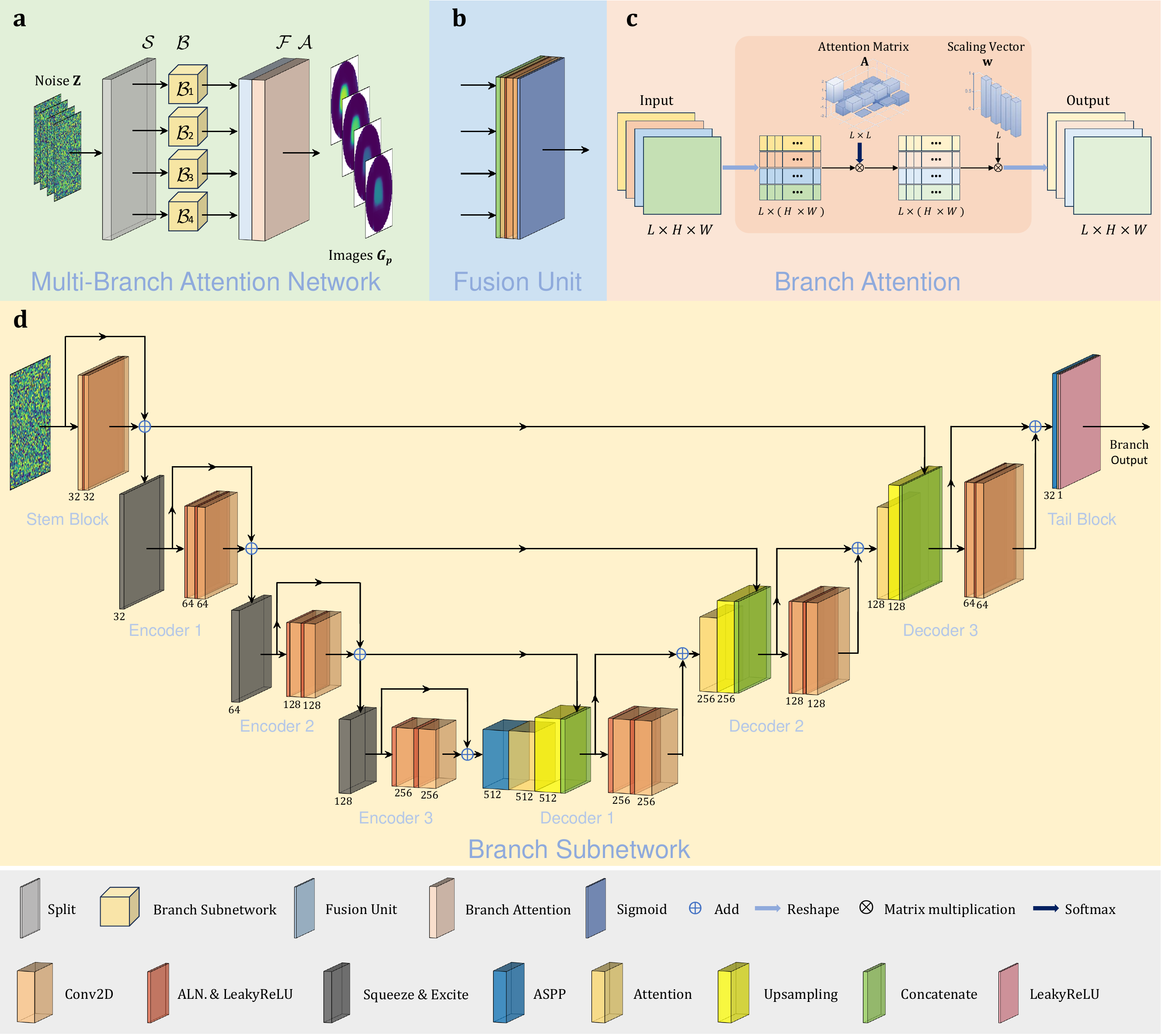}
\caption{The architecture of MBA-Net: a) the overall architecture; b) the Fusion Unit, used for multi-branch feature fusion; c) the structure of the Branch Attention module, detailing how attention mechanisms are applied, with the attention matrix $\mathbf{A}$ and the scaling vector $\mathbf{w}$ illustrating examples of their respective final values; and d) the architecture of the branch subnetwork, showing the configuration and connectivity of modules of our branch subnetwork.}
\label{network}
\end{figure*}

We employ the $Adam$  to solve (\ref{prob}), a stochastic gradient descent method known for its fast convergence and robustness to noise in various optimization problems \cite{b36}. We opt for the $\ell_1$ norm over the Frobenius norm due to the superior structure preservation it offers in mfEIT reconstruction (see Fig.~\ref{ablation}). Suppose $\mathcal{T} (\pmb{\theta}|\mathbf{Z}) = \|\mathbf{\widetilde{J}} \mathcal{R}_{v}(\phi(\pmb{\theta}|\mathbf{Z}))-\mathbf{V}\|_1$, the first step in the $Adam$ optimization framework is to calculate the gradient of $\mathcal{T}$ with respect to $\pmb{\theta}$, i.e.:
% To use Adam, we firstly suppose the l1 norm $\mathcal{T} (\pmb{\theta}|\mathbf{Z}) = \|\mathbf{\widetilde{J}} \mathcal{R}_{v}(\phi(\pmb{\theta}|\mathbf{Z}))-\mathbf{V}\|_1$ and  then calculate the gradient of $\mathcal{T}$ with respect to the $\pmb{\theta}$, i.e.:
\begin{equation} \label{grad}
\frac{\partial \mathcal{T}}{\partial \pmb{\boldsymbol{\theta}}} =   \left(\frac{\partial \mathcal{T}}{\partial \phi(\pmb{\theta}|\mathbf{Z})} \right)^T \frac{\partial \phi(\pmb{\theta}|\mathbf{Z})}{\partial \pmb{\boldsymbol{\theta}}}.
\end{equation}

The gradient (\ref{grad}) is calculated using $PyTorch$'s built-in automatic differentiation engine called $torch.autograd$, since each term in (\ref{prob}) is expressed by tensors. Additionally, as the $\ell_1$ norm is non-differentiable, $\partial \mathcal{T}/\partial \phi(\pmb{\theta}|\mathbf{Z})$ is approximated by its subgradient. The parameter $\pmb{\theta}$ is then updated according to the $Adam$ parameter update rules. The final multi-frequency conductivity distribution $\Hat{\pmb{\Sigma}}$ can be obtained by:
\begin{equation}
\Hat{\pmb{\Sigma}}=\mathcal{R}_{v}\left(\phi\left(\pmb{\theta}^{(t)}|\mathbf{Z}\right)\right),
\end{equation}
where $t \in \mathbb{N}_{+}$ stands for the number of iterations and is treated as a parameter in the MAIP method. Another parameter in the iteration stage is the learning rate in $Adam$ (represented by $lr \in \mathbb{R}_{+}$). The schematic of the MAIP algorithm is shown in Fig.~\ref{MAIP}.

% The algorithm begins by specifying the required inputs, including the number of iterations $\mathcal{N}$, the learning rate \mathcal{lr}, the initial network parameters $\pmb{\theta}^{(0)}$, the voltage measurements $\mathbf{V}$, the normalized and mapped sensitivity matrix $\mathbf{\widetilde{J}}$, and the input noise matrix $\mathbf{Z}$. 

% The Adam optimizer is then executed for a total of \mathcal{N} iterations, during which the network parameters are iteratively updated to minimize the 
% L1 norm of the difference between the predicted voltage measurements and actual measurements. This optimization process is expressed as: 

% \begin{equation}
% \pmb{\theta}^{(k)}=\arg\min_{\mathbf{\pmb{\theta}}}\|\mathbf{\widetilde{J}}f(\pmb{\theta}^{(k-1)}|\mathbf{Z})-\mathbf{V}\|_{1},
% \end{equation}
% where $\pmb{\theta}^{(k)}$ represents the network parameters at the $k\text{-th}$ iteration.

% Upon completing the specified number of iterations, the final reconstructed multifrequency conductivity distribution matrix $\hat{\mathbf{\Sigma}}$ is obtained as the output of the neural network. Here, we can rewrite \eqref{e7} as: 
% \begin{equation}
% \hat{\mathbf{\Sigma}}=f(\pmb{\theta}^{(\mathcal{N})}|\mathbf{Z}),
% \end{equation}
% where $\mathcal{N}$ denotes the total number of iterations of the Adam optimizer.

\subsection{Network Architecture}
The MBA-Net (see Fig.~\ref{network}a) adopted in the MAIP framework consists of three components: a set of multiple branch subnetworks, denoted by $\mathcal{B}: \wedge^L ~ \mathbb{R}^{1\times H\times W}\to \wedge^L~ \mathbb{R}^{1\times H\times W}$; a fusion unit (FU), represented by $\mathcal{F}: \wedge^L ~ \mathbb{R}^{1\times H\times W}\to \mathbb{R}^{L\times H\times W}$; and a branch attention (BA) module, expressed as $\mathcal{A}:\mathbb{R}^{L\times H\times W} \to \mathbb{R}^{L\times H\times W}$. Here, $\wedge^L$ represents the $L$-fold Cartesian product.

% The input for the MBA-Net is a random noise tensor $\mathbf{Z}\in\mathbb{R}^{L\times H\times W}$, and the output of the network is the mfEIT reconstructed images $\boldsymbol{G}_{p}\in\mathbb{R}^{L\times H\times W}$. 
% The total number of parameters for the proposed MBA-Net is approximately $5.79\times10^7$.
After being fed into the MBA-Net, the multi-channel input $\mathbf{Z}\in\mathbb{R}^{L\times H\times W}$ is decomposited into $L$ single-channel noise images, denoted by $\mathbf{z}_{i} \in\mathbb{R}^{1\times H\times W}$ ($i=1,2,...,L$), each of which is then passed into a corresponding branch subnetwork, i.e.:
\begin{equation}
    \left\{\mathbf{B}_{i}\right\}_{i=1}^L=\left\{\mathcal{B}_i(\mathbf{z}_i)\right\}_{i=1}^L = \mathcal{B}\left( \{\mathbf{z}_i\}_{i=1}^L \right),
\end{equation}
where $\mathcal{B}_i: \mathbb{R}^{1\times H\times W} \to \mathbb{R}^{1\times H\times W}$ represents the $i\text{-th}$ branch subnetwork, and $\mathbf{B}_{i}$ stands for its output.

Next, $\{\mathbf{B}_{i}\}_{i=1}^L$ passes through the FU (Fig.~\ref{network}b), which integrates the multi-branch information into a unified feature map $\mathbf{F}\in\mathbb{R}^{L\times H \times W}$. $\mathbf{F}$ is then fed into the meticulously designed BA module, which refines the fused features through a channel-wise attention mechanism by selectively emphasizing the cross-channel salient features through a learnable attention matrix. Simultaneously, it improves the conductivity contrasts in the reconstructed mfEIT images across different channels using a learnable scaling vector. The introduction of the BA module not only enhances the robustness and accuracy of mfEIT image reconstruction (see Fig.~\ref{ablation} and Table.~\ref{table_ablation}), but also enhances the interpretability of the network. For instance, by examining the ultimate entries of the attention matrix (see Fig.~\ref{network}c), we can identify which frequency-specific features are prioritized, thus providing insights into how different frequency measurements contribute to the final reconstructed images.
%In the BA module, attention scores are computed from a learnable weight matrix to assess the importance of each channel. These scores weight the feature maps via batch matrix multiplication, enabling the network to emphasize the most relevant features across different channels while suppressing less significant ones, such as measurement noise or low-impact frequency data. Additionally, each channel is scaled by learnable weights, offering a secondary level of refinement, particularly enhancing inter-channel differences. This dual mechanism—channel attention and weighted scaling—enhances the accuracy and robustness of the reconstructed images. 
% Finally, the output from the BA module is split into $L$ single-channel $1\times H\times W$ images, which are mfEIT reconstructed images. 
Suppose we denote the split operation as $\mathcal{S}: \mathbb{R}^{L\times H\times W}\to \wedge^L ~\mathbb{R}^{1\times H\times W}$, the reconstructed mfEIT images $\boldsymbol{G}_{p}\in\mathbb{R}^{L\times H \times W}$ can be expanded as:
\begin{equation}\boldsymbol{G}_{p}=\phi\left(\pmb{\theta}^{(t)}|\mathbf{Z}\right)= \mathcal{A}\left( \mathcal{F}\left( \mathcal{B}\left( \mathcal{S}\left( \mathbf{Z} \right)\right)\right)\right).
\end{equation}

We describe key modules of MBA-Net in the following parts.

\subsubsection{Branch Subnetwork}
Previous studies using U-Net-like architectures have shown superior performance in DIP-based tomographic tasks \cite{b18,b24,b25}. Here, we design a ResUNet as the branch subnetwork, inspired by ResUNet++\cite{b28}. Compared to the U-Net used in DeepEIT\cite{b18}, our residual design and Adaptive Layer Normalization (ALN) improve the convergence ability, providing stable convergence performance without the need for specially designed stopping criteria\cite{b18} (see Fig.~\ref{loss_noise_curve1}). Additionally, we use LeakyReLU to avoid the dying neuron problem. The architecture of the branch subnetwork is illustrated in Fig.~\ref{network}d.

The input for each branch subnetwork is a single-channel noise image, i.e. $\mathbf{z}_i$. Each branch subnetwork begins with a stem block, followed sequentially by three encoder blocks, an Atrous Spatial Pyramid Pooling (ASPP) module\cite{b43}, and then three decoder blocks, finally ending with a tail block. The stem block consists of an Adaptive Layer Normalization (ALN) module, a Leaky Rectified Linear Unit (LeakyReLU), and two convolutional layers. %ALN stabilizes the network iteration process by normalizing all features within each sample. Its details will be provided in the next subsection. LeakyReLU addresses the dying ReLU problem by allowing a small positive slope for negative inputs, thereby increasing the model's nonlinearity. 
All negative slopes in LeakyReLU are set to 0.0001 based on trial and error. In comparison to the stem block, each encoder block additionally includes an Squeeze-and-Excitation (SE) block\cite{b42}, an ALN module and a LeakyReLU. Each encoder's first convolutional layer is strided to reduce the spatial dimensions of the feature maps by half. %In order to enhance the representational power of the network, This allows the network to adaptively amplify important features and suppress less relevant ones. 
ASPP serves as a bridge, expanding the filters' receptive field to encompass a more extensive context. Each decoder block, compared to the encoder, replaces the SE block with an attention module and an upsampling module. Within it, the attention module is applied to the feature maps to enhance their expressive power, followed by upsampling using bilinear interpolation to restore the spatial resolution reduced by strided convolution in the encoder, and concatenation operation to integrate features from the corresponding encoding path. The tail block, which refines the final output, consists of an ASPP module, followed by a $1\times 1$ convolution and a LeakyReLU activation.

\begin{figure}
\centerline{\includegraphics[scale=0.36]{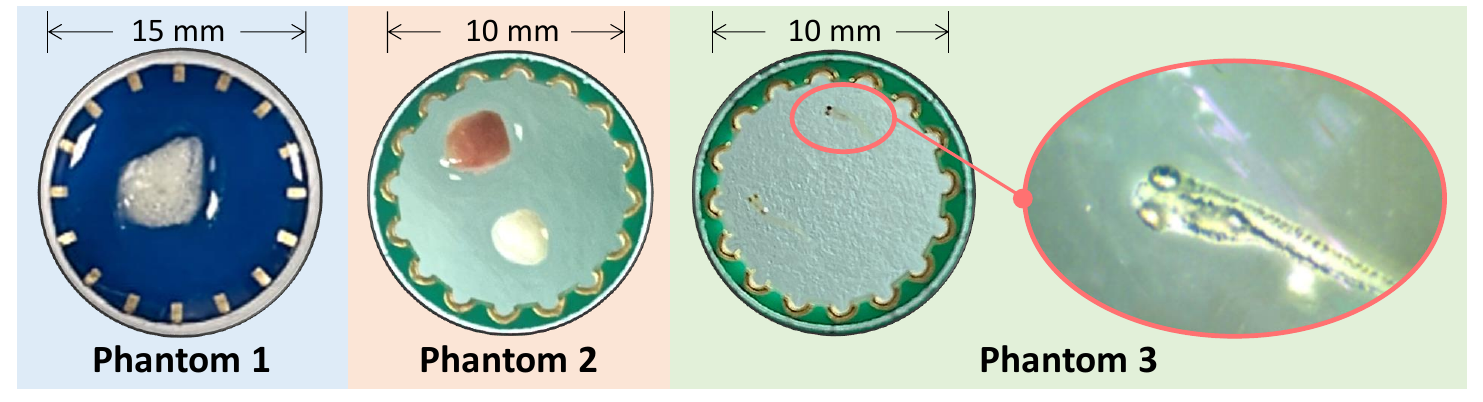}}
	\caption{Real-world experiment phantoms: Phantom 1: apple flesh in a 15 mm 16-electrode EIT sensor; Phantom 2: sheep liver (top) and chicken skin slices (down) in a 10 mm 16-electrode EIT sensor; Phantom 3: two zebrafishes within the same sensor as phantom 2.}
	\label{real-world_exp}
\end{figure}

\subsubsection{Adaptive Layer Normalization}
The MAIP algorithm is a training-data-free approach where the batch size is always set to $1$. This leads to  Batch Normalization \cite{b29,b30} being ineffective in our case. To address this, we replace BN with Layer Normalization (LN) \cite{b31} in the convolutional layers.

LN stabilizes training by normalizing features within each sample, making it particularly effective when the batch size is 1. Unlike BN that normalizes over the batch dimension,  LN normalizes across all dimensions except the last one of the input tensor. To ensure LN works effectively on input feature maps of varying sizes, we developed an adaptable LN module: Given a feature tensor with $c$ channels and $h \times w$ spatial dimensions, we first reshape the tensor from $c\times h\times w$ to $(h \times w)\times c$. After reshaping, LN is applied by computing the mean $\mu$ and standard deviation $\delta$ across all channels, which are then used to normalize the 2D tensor along channel axis. Finally, the normalized tensor is reshaped back to its original form. The normalization process is formulated as follows:
\begin{equation}
\mu=\frac1c\sum_{j=1}^c\mathbf{x}_j,
\label{e9}
\end{equation}
\begin{equation}
\delta=\sqrt{\frac{1}{c}\sum_{j=1}^c(\mathbf{x}_j-\mu)^2},
\label{e10}
\end{equation}
\begin{equation}
\mathbf{y}_j=\frac{\mathbf{x}_j-\mu}{\delta+\epsilon},
\label{e11}
\end{equation}
where $\mathbf{x}_j$ is the vector of the $j\text{-th}$ channel after reshaping and $\mathbf{y}_j$ is its normalized version. $\epsilon=1\times10^{-5}$ is a small positive number added for numerical stability. Note that operations in \eqref{e10} and \eqref{e11} are element-wise across the vectors.

\subsubsection{Fusion Unit}
The FU (Fig.~\ref{network}b) aims to initially fuse the multi-branch information. The output of the FU, represented as a tensor $\mathbf{F}\in\mathbb{R}^{L\times H \times W}$, can be readily expressed as: %It integrates the outputs of the four branch subnetworks by first concatenating them along the channel dimension. This combined feature map is then processed through two 3 × 3 convolutional layers, followed by a 1×1 convolution. Finally, a Sigmoid activation is applied to normalize the output to a range of 0 to 1. 
\begin{align}
    \mathbf{F} & =  \mathcal{F}\left(\left\{\mathbf{B}_{i}\right\}_{i=1}^L\right) \nonumber\\ 
    &=\mathrm{Sigmoid}((\mathbf{W}_{1}*(\overline{\mathbf{W}}_{3}*(\mathbf{W}_{3}*(\oplus_{i=1}^L\mathbf{B}_i))))),
\end{align}
where $\oplus$ denotes concatenation operation and $*$ represents the convolution. $\mathbf{W}_{3}$ and $\overline{\mathbf{W}}_{3}$ are the weight matrices for the first and second 3 × 3 convolutional layers, respectively. $\mathbf{W}_{1}$ is the weight matrix for the 1 × 1 convolutional layer. 
$\mathrm{Sigmoid}(\cdot)$ represents the $\mathrm{Sigmoid}$ function.

\subsubsection{Branch Attention}
The BA module is designed to further integrate and utilize the information from FU. 
%by selectively emphasizing the most important features across these channels through a learnable attention matrix. Simultaneously, it captures the conductivity differences in the mfEIT reconstructed images across different channels using a learnable scaling vector. This attention mechanism not only enhances feature integration but also improves the interpretability of the network. 
The overall structure of the BA module is illustrated in Fig.~\ref{network}c, whose input is $\mathbf{F}$. We define a learnable channel attention weight matrix as $\mathbf{A}\in\mathbb{R}^{L\times L}$ and a learnable channel scaling vector as $\mathbf{w}\in\mathbb{R}^{L}$. $\mathbf{A}$ is randomly initialized along with the network parameters, while $\mathbf{w}$ is initialized with all elements set to 1. The workflow of the BA module is outlined as follows.

First, a $\mathrm{Softmax}$ function is applied to each row of $\mathbf{A}$:
\begin{equation}
\bar{\mathbf{A}} = \mathrm{Softmax}(\mathbf{A}),
\end{equation}
where $\bar{\mathbf{A}}$ denotes the normalized attention matrix.

Subsequently, the input of the BA module, i.e. $\mathbf{F}$, is reshaped into a two-dimensional feature matrix $\mathbf{F}^{\prime}\in\mathbb{R}^{L\times (H \times W)}$ by:
\begin{equation}
\mathbf{F}'=\mathcal{R}_{1}(\mathbf{F}),
\end{equation}
where $\mathcal{R}_{1}:\mathbb{R}^{L\times H\times W}\to\mathbb{R}^{L\times(H\times W)}$ denotes the reshaping operation. $\bar{\mathbf{A}}$ is then applied to the $\mathbf{F}^{\prime}$ via matrix multiplication followed by another matrix multiplication with  $\mathbf{w}$. Finally, the output of the BA module, exactly the mfEIT images $\boldsymbol{G}_{p}\in\mathbb{R}^{L\times H \times W}$, is obtained by reshaping $\mathbf{w}\bar{\mathbf{A}}\mathbf{F}'$ back to the its input dimension using another tensor reshaping operation $\mathcal{R}_{2}:\mathbb{R}^{L\times (H\times W)}\to\mathbb{R}^{L\times H\times W}$. Consequently, $\boldsymbol{G}_{p}$ is expressed as:
\begin{equation}
\boldsymbol{G}_{p} = \mathcal{A} \left(\mathbf{F} \right)= \mathcal{R}_{2}(\mathbf{w}\bar{\mathbf{A}}\mathcal{R}_{1}(\mathbf{F})).
\end{equation}

% As the BA module is placed at the end of the MBA-Net in this work, it should be noted that $\boldsymbol{G} = \phi\left(\pmb{\theta}|\mathbf{Z}\right)$.

\section{Experimental Setup}
\subsection{Simulation Data}
We obtained simulation data using COMSOL Multiphysics with a modeled 16-electrode circular EIT sensor by placing phantoms of various shapes, quantities and conductivity. The background medium and electrodes were set to be physiological saline and titanium, respectively, with conductivities of $2$ S/m and $7.407\times10^5$ S/m. Phantoms of different shapes, including circles, rectangles, and triangles, were placed within the imaging area, with their conductivities gradually changing to simulate the frequency-dependent characteristics of tissues. 

\begin{figure}[!t]
	\centerline{\includegraphics[scale=0.24]{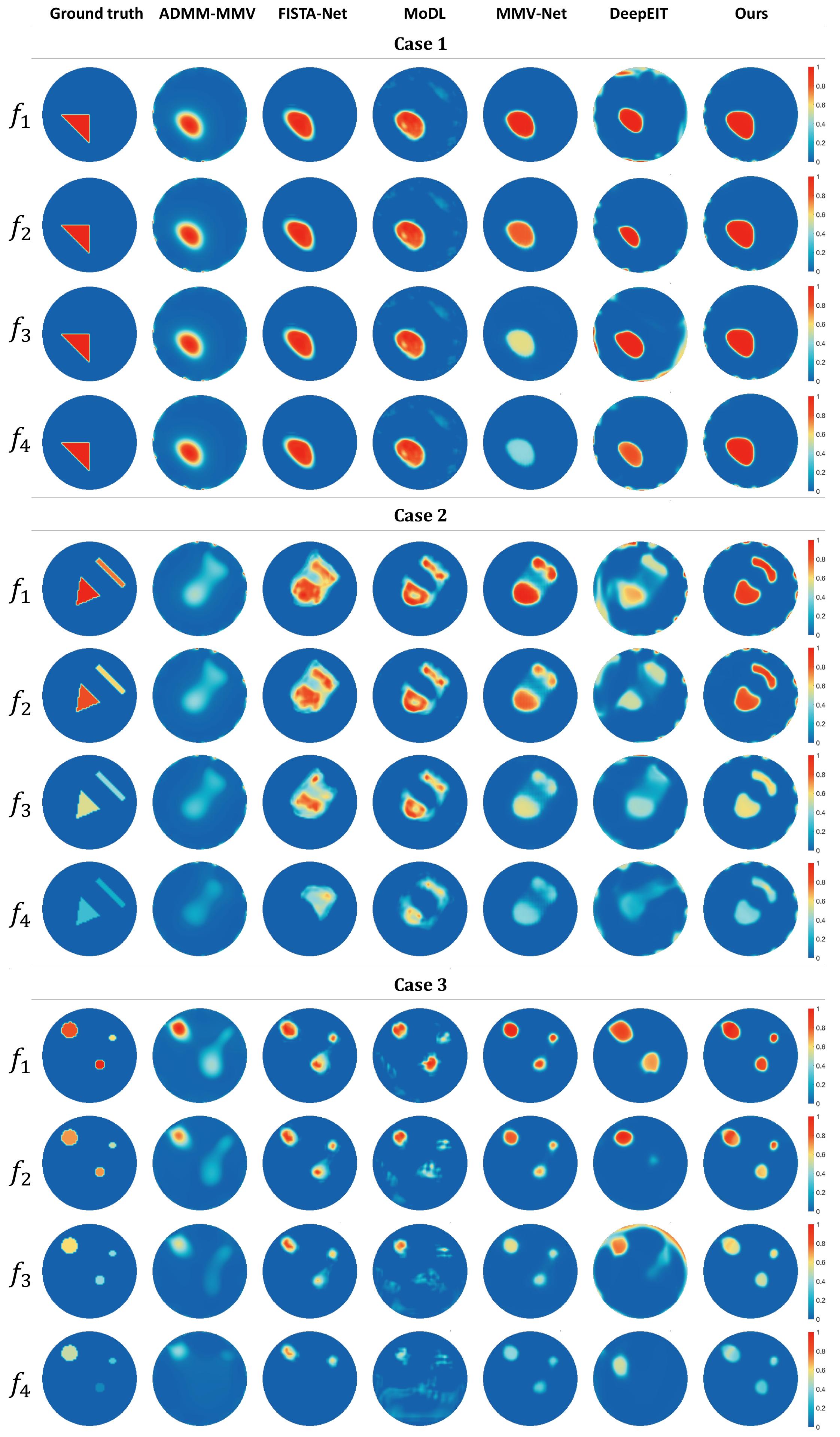}}
	\caption{Comparative results of the proposed MAIP algorithm and SOTA image reconstruction algorithms (i.e., ADMM-MMV\cite{b7}, FISTA-Net\cite{b20}, MoDL\cite{b19}, MMV-Net\cite{b9}, and DeepEIT\cite{b18}) in three simulated cases. }
	\label{simulation_results}
\end{figure}

We designed three simulation experiments (as illustrated on the left of Fig.~\ref{simulation_results}), applying FD-mfEIT for Case 1 and Case 2, and TD-mfEIT for Case 3. In Case 1, in addition to the reference measurement, we performed a single measurement and replicated it four times to investigate the ability of different mfEIT reconstruction algorithms to accurately capture inter-frequency correlations. Case 2 was designed to evaluate the performance of different algorithms in reconstructing multiple complex shapes. Given the ill-posed nature of the EIT image reconstruction problem, accurately reconstructing shapes such as triangles and rectangles is challenging. Case 3, on the other hand, was designed to further assess the performance of various algorithms in reconstructing multiple objects, particularly when the imaging targets vary substantially in size and conductivity.

\begin{table}
\caption{Quantitative Comparisons (RIE, CC, PSNR, MSSIM and PA-MSSIM) Based on Simulation Results.}
\centering
\includegraphics[scale=0.335]{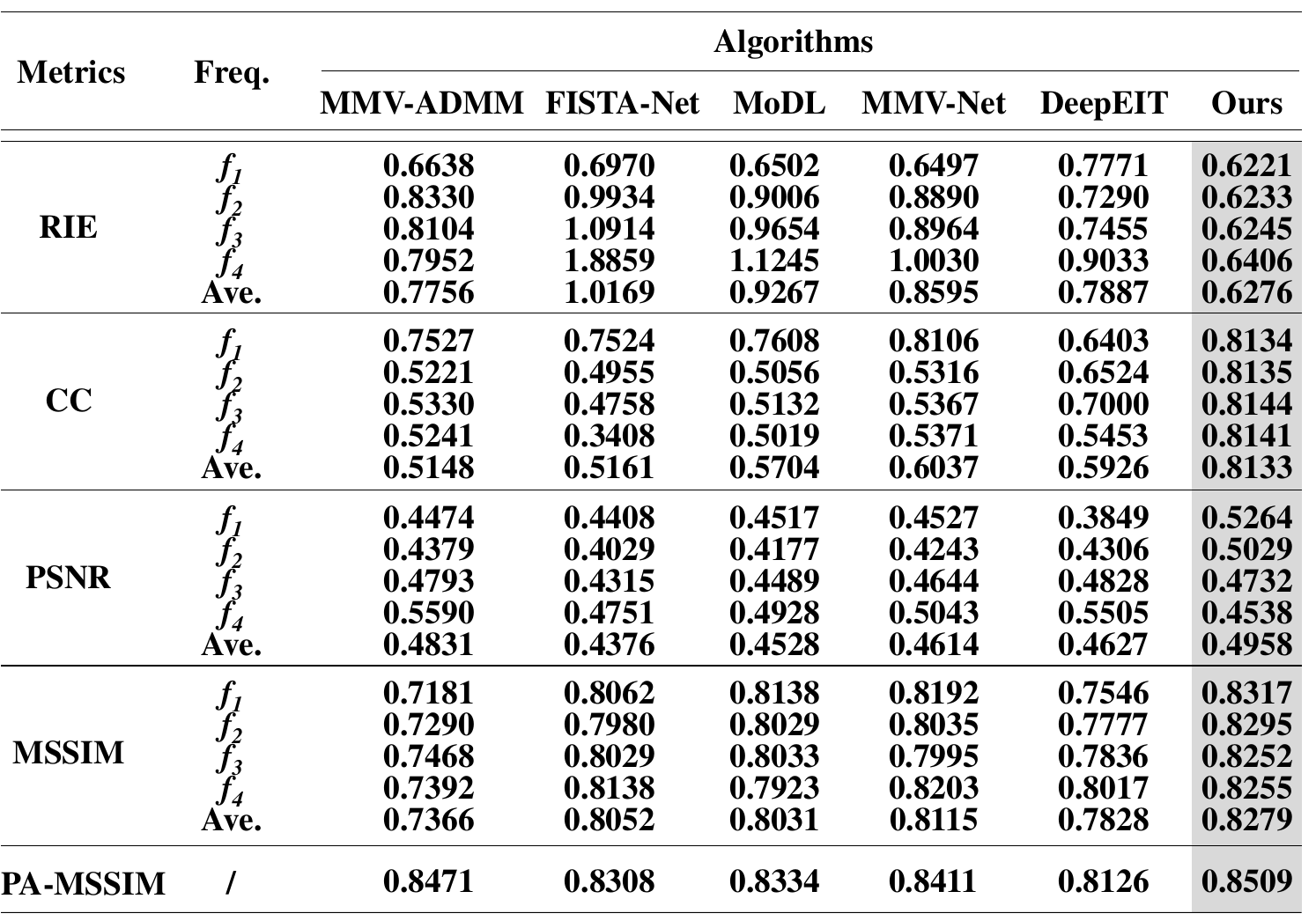}
\label{metrics_1}
\vspace{0.3cm}
\end{table}

\begin{figure}[!t]
	\centerline{\includegraphics[scale=0.45]{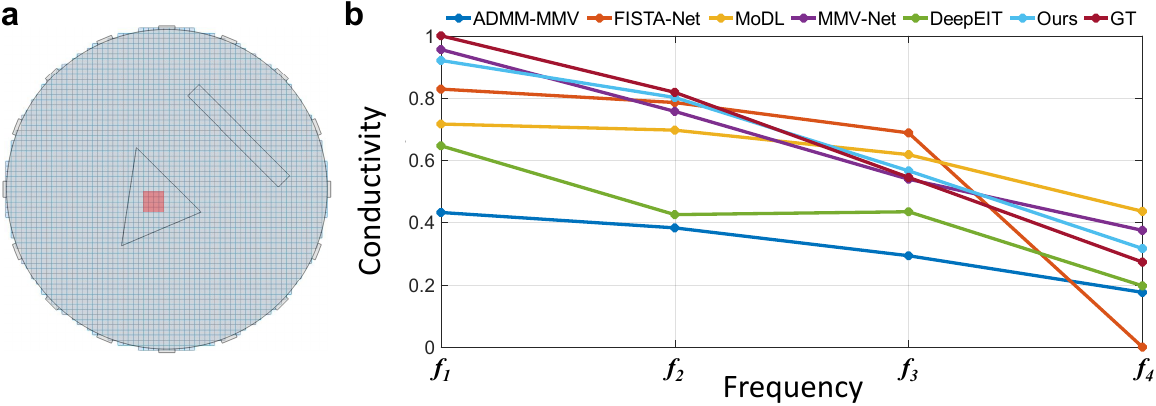}}
	\caption{Comparison of average conductivity curves in the selected region: (a) the selected region (4x4 pixels highlighted in red); (b) ground truth and reconstructed images' average conductivity curves in the selected region. }
	\label{GtvsPred}
 % \vspace{0.3cm}
\end{figure}

% The conductivity distributions across different frequencies are denoted by $\mathbf{\sigma} = [\mathbf{\sigma}_0, \mathbf{\sigma}_1, \mathbf{\sigma}_2, \mathbf{\sigma}_3, \mathbf{\sigma}_4]$, where each $\sigma_j$ represents the conductivity at a specific frequency. Five sets of measurements were conducted using adjacent measurement protocol, and the corresponding voltage data, $\mathbf{v} = [\mathbf{v}_0, \mathbf{v}_1, \mathbf{v}_2, \mathbf{v}_3, \mathbf{v}_4]$, were recorded. We set $\mathbf{v}_0$ as the reference voltage and normalized the remaining measurements against it, ultimately obtaining the simulated voltage data for mfEIT image reconstruction. 
% \begin{equation}
% \Delta\mathbf{v}_{j}=-(\mathbf{v}_{j}-\mathbf{v}_{0})\oslash \mathbf{v}_{0} 
% \end{equation}
% \begin{equation}
% \Delta\pmb{\sigma}_{j}=(\pmb{\sigma}_{j}-\pmb{\sigma}_{0})\oslash \pmb{\sigma}_{0} 
% \end{equation}
% where $\oslash$ represents the Hadamard division. $j=1,2,3,4$ represents the $j\text{-} th$ simulation, and $\Delta\mathbf{\sigma}_{j}$ gives the ground truth of the $j\text{-} th$ simulation.

\subsection{Real-world Data}
We designed three sets of phantom experiments to obtain real-world data, as illustrated in Fig.~\ref{real-world_exp}. The first set of experiments was conducted using a miniature EIT sensor with an inner diameter of 15 mm and 16 planar electrodes. The phantom tissue used in the experiment is fresh apple flesh. FD-mfEIT is applied for this experiment and the excitation current frequencies were 100 kHz, 50 kHz, 40 kHz, 20 kHz, and 10 kHz, with 10 kHz serving as the reference frequency.

The second and third sets of experiments employed a 10 mm miniature EIT sensor with 16 electrodes. The phantom consisted of animal tissue slices, including sheep liver and chicken skin. In the third set, the imaging targets were two four-day-old zebrafish larvae, a species widely used in oncology and genetics research. The larvae were anaesthetised using MS222 during the imaging process, and they are not regulated by the Home Office as protected animals. Both the second and third experiments utilized the TD-mfEIT approach. The excitation current frequencies were 10 kHz, 20 kHz, 50 kHz, and 70 kHz, with measurements taken in the background medium alone used as the reference. For all real-world experiments, we use saline with a conductivity of approximately 0.07 S/m as the background medium. Within our investigated frequency range, the conductivity of saline can be considered frequency-independent\cite{b46}.

\subsection{Comparison Algorithms}
We compared the performance of our MAIP algorithm against five SOTA tomographic imaging algorithms: ADMM-MMV\cite{b7}, MoDL\cite{b19}, FISTA-Net\cite{b20}, MMV-Net\cite{b9}, and DeepEIT\cite{b18}. Among these algorithms, FISTA-Net and MoDL are model-based supervised learning SMV algorithms and DeepEIT is an unsupervised learning SMV algorithm.
ADMM-MMV is a traditional model-based MMV algorithm, as well as MMV-Net belongs to model-based supervised learning MMV methods. All supervised learning methods used in our experiments were trained on the $\textit{Edinburgh mfEIT Dataset}$\cite{b9} using $PyTorch$ and the $Adam$ optimizer.

\begin{figure}[!t]	\centerline{\includegraphics[scale=0.43]{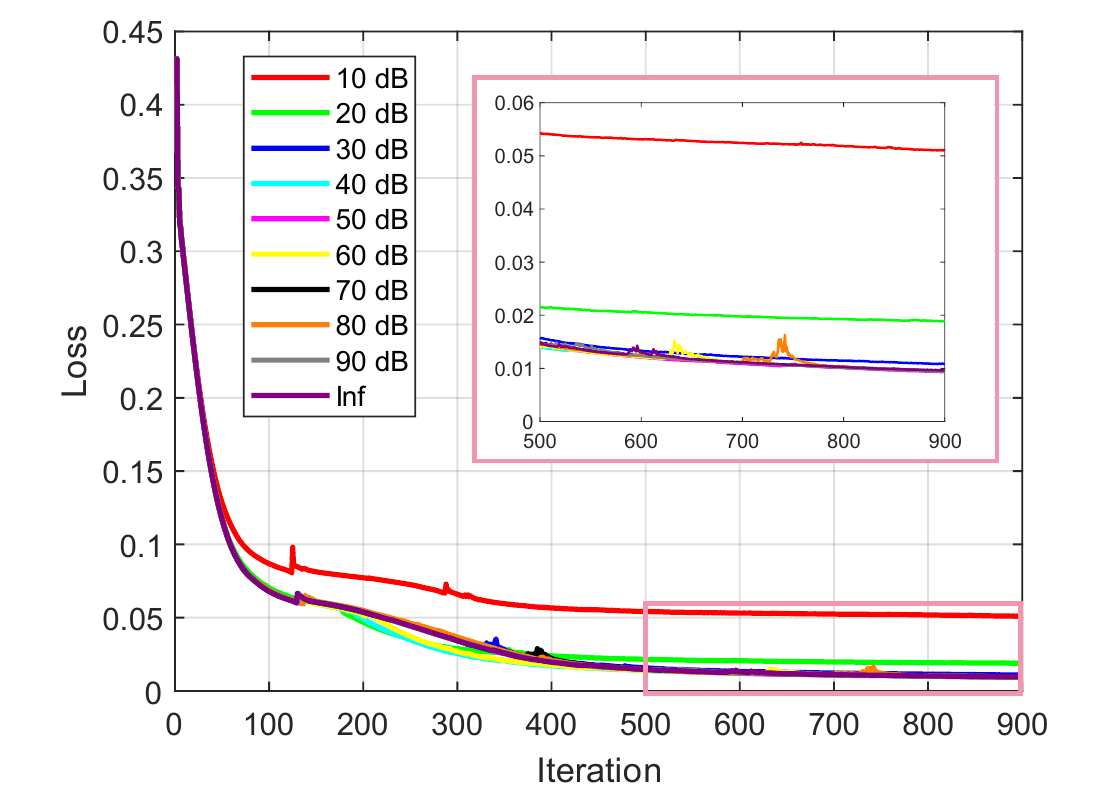}}
	\caption{Simulation results of Case 2: loss curves of MAIP under different noise levels.}
	\label{loss_noise_curve1}
\end{figure}

\subsection{Parameter Settings}
In addition to $t$ and $lr$, the parameters for our MAIP algorithm include network initialization parameters. We employ Kaiming initialization\cite{b44} for the MBA-Net and fix the initial network parameters through trial and error based on extensive experimentation. In all experiments, $t$ is set to 900, and $lr$ is fixed at 0.00012. For comparison, the number of iterations and learning rate for DeepEIT are set to 8000 and 0.005 in simulations, and 8000 and 0.001 in real-world experiments. The parameters of DeepEIT were also carefully tuned based on extensive experiments to ensure a fair comparison.

\begin{figure}[!t]	\centerline{\includegraphics[scale=0.46]{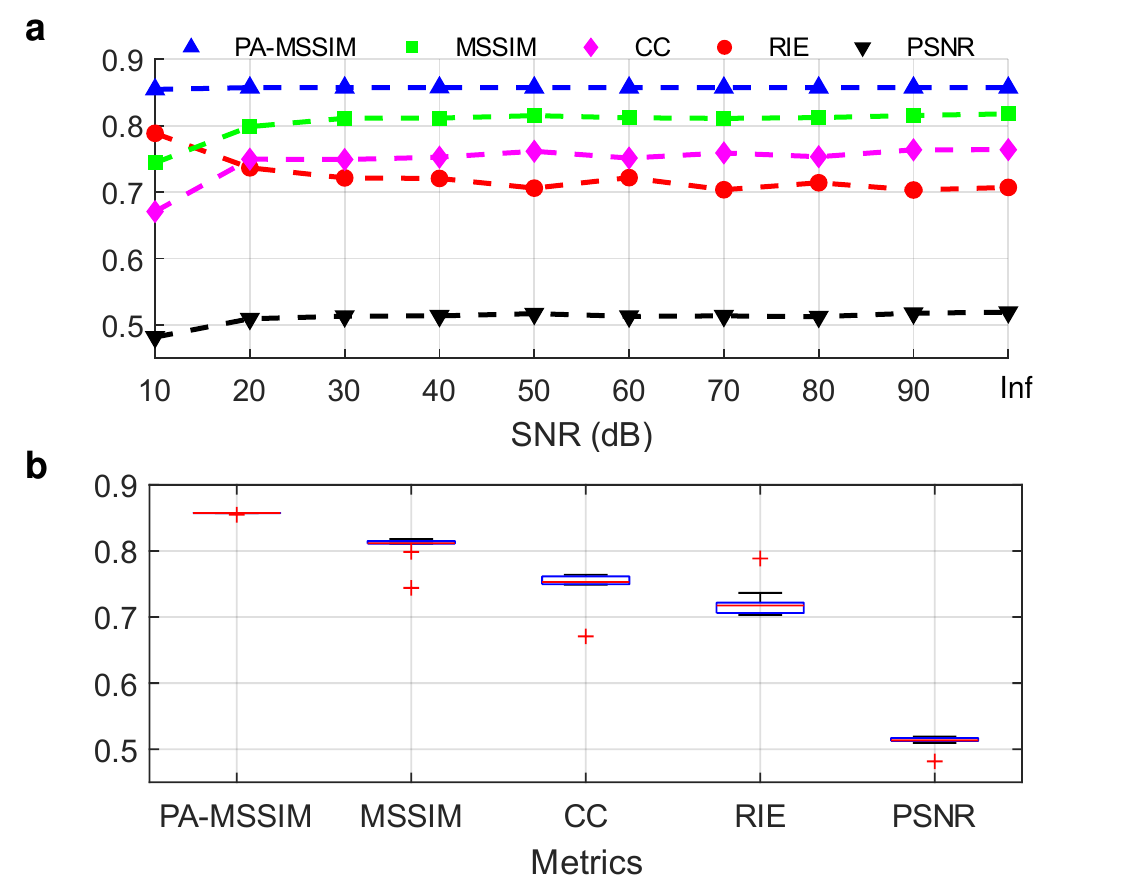}}
	\caption{Simulation results of Case 2: (a) effect of noise on quantitative metrics; (b) the boxplot of quantitative metrics.}
	\label{noise_resistance}
\end{figure}

\begin{figure}[!t]
	\centerline{\includegraphics[scale=0.25]{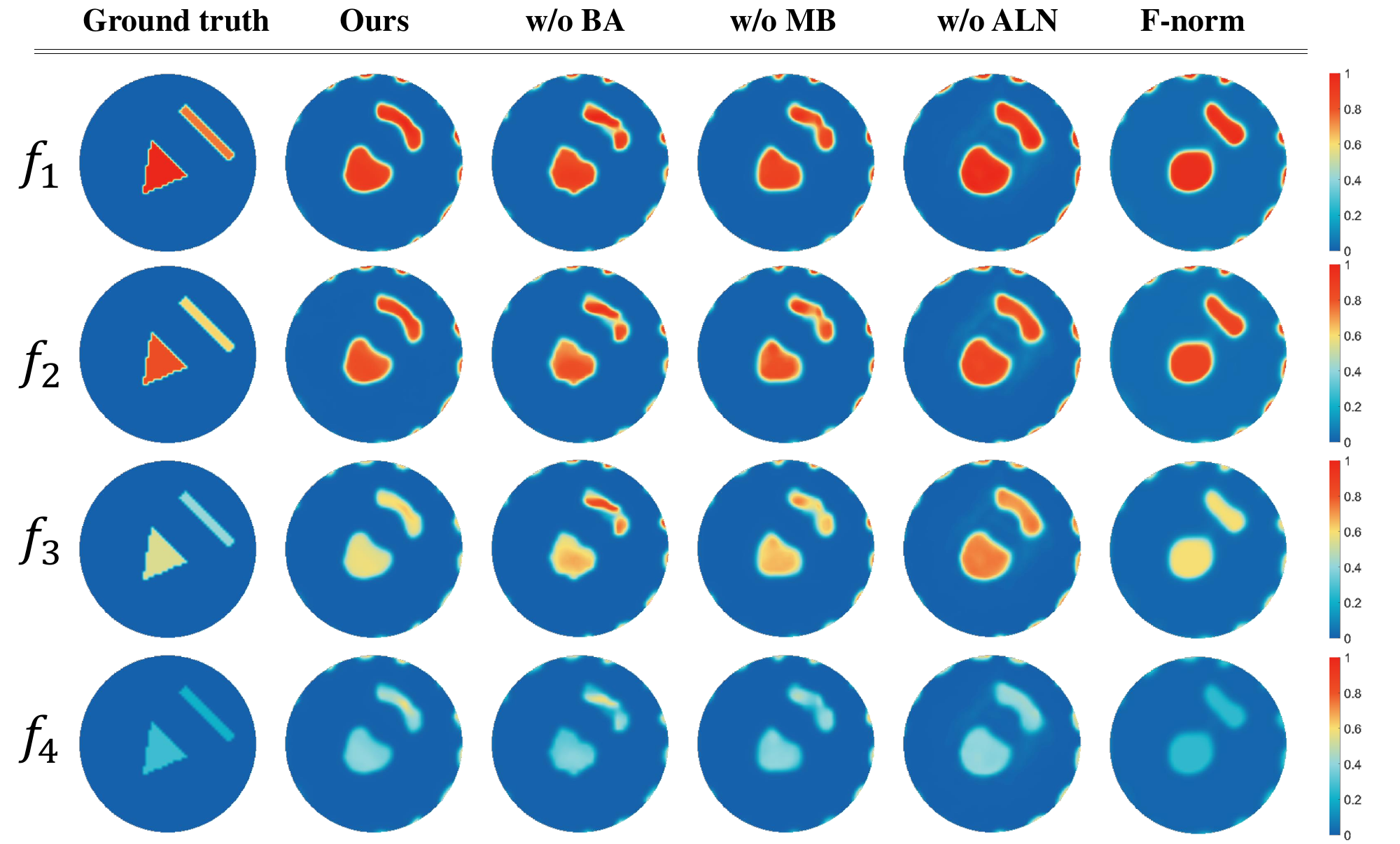}}
	\caption{Ablation study of the proposed MAIP algorithm on Case 2. }
	\label{ablation}
\end{figure}

\subsection{Quantitative Metrics}
We normalize the ground truth and reconstructed results using maximum value normalization and compare them using four metrics: Relative Image Error (RIE), Correlation Coefficient (CC), Peak Signal-to-Noise Ratio (PSNR), and Mean Structural Similarity Index Measure (MSSIM). RIE measures the overall difference between the reconstructed image and the reference image, thereby quantifying pixel-level accuracy. CC evaluates the linear correlation between the pixel intensities of the reconstructed image and the ground truth. PSNR quantifies how closely the reconstructed image matches the ground truth in terms of pixel intensity, taking noise into account. For easier visualization, we use scaled PSNR values. MSSIM assesses the structural similarity between the reconstructed image and the reference image, reflecting how well the structural information is preserved. 
Assume that a frame of the reconstructed image and the ground truth are denoted as $\boldsymbol{g}_p\in\mathbb{R}^{H \times W}$ and $\boldsymbol{g}_g\in\mathbb{R}^{H \times W}$, respectively. These metrics are defined as:
\begin{equation}
\mathrm{RIE}=\frac{||\boldsymbol{g}_p-\boldsymbol{g}_g||_{F}}{||\boldsymbol{g}_g||_{F}},
\end{equation}
\begin{equation}
    \mathrm{CC}=\frac{(\boldsymbol{g}_p-\bar{g}_{p})\cdot(\boldsymbol{g}_g-\bar{g}_{g})}{\|\boldsymbol{g}_p-\bar{g}_{p}\|_F\cdot\|\boldsymbol{g}_g-\bar{g}_{g}\|_F},
\end{equation}
\begin{equation}  \mathrm{PSNR}=\frac{1}{4}  \cdot\log_{10}\left(\frac{H \times W}{\|\boldsymbol{g}_p-\boldsymbol{g}_g\|_F^2}\right),
\end{equation}
\begin{equation}
\mathrm{MSSIM}=\frac1{N_{b}}\sum_{q=1}^{N_{b}}\frac{\left(2{\kappa}_{p,q}{\kappa}_{g,q}+C_1\right)\left(2{\tau}_{pg,q}+C_2\right)}{\left({\kappa}_{p,q}^2+{\kappa}_{g,q}^2+C_1\right)\left({\tau}_{p,q}^2+{\tau}_{g,q}^2+C_2\right)},\end{equation}
where $\| \cdot \|_F$ denotes the Frobenius norm. $\bar{g}_{p}$ and $\bar{g}_{g}$ are the average values of the pixels in $\boldsymbol{g}_p$ and $\boldsymbol{g}_g$, respectively. $N_{b}$ is the number of block pairs, and $q=1,2,...,N_{b}$ represents the $q\text{-th}$ pair of corresponding blocks from images $\boldsymbol{g}_p$ and $\boldsymbol{g}_g$ used for calculating the SSIM index. ${\kappa}_{p,q}$, ${\kappa}_{g,q}$, ${\tau}_{p,q}$, ${\tau}_{g,q}$, and ${\tau}_{pg,q}$ represent the local means, standard deviations, and covariance of these block pairs, respectively. Furthermore, in MSSIM calculation, the standard deviation of the isotropic Gaussian function is set to 0.2, while the scalar constants for luminance $C_1$, contrast $C_2$, and structural terms $C_3$ are set as 0.0001, 0.0009 and 0.00045, respectively.

Additionally, to compare the ability of different algorithms to maintain inter-frequency structural consistency in mfEIT, we introduce the Pairwise Average MSSIM (PA-MMSIM) to calculate the average of the MSSIM values between pairs of conductivity images at different frequencies. Given an mfEIT reconstruction task with $L$ observed frequencies, then for the $L$ reconstructed conductivity images $\boldsymbol{g}_{p,1}$, $\boldsymbol{g}_{p,2}$, $...$, $\boldsymbol{g}_{p,L}$, the calculation of the PA-MSSIM can be defined as follows:
\begin{equation}
\mathrm{PA\text{-}MSSIM}=\frac{1}{\binom{L}{2}}\sum_{m=1}^{L-1}\sum_{n=m+1}^{L}\mathrm{MSSIM}(\boldsymbol{g}_{p,m},\boldsymbol{g}_{p,n}),
\end{equation}
here $\binom{L}{2}=\frac{L(L-1)}2$ represents the combinational number.

\section{Results And Discussion}

\begin{table}[!t]
\caption{Quantitative Comparisons (RIE, CC, PSNR, MSSIM and PA-MSSIM) for Ablation Study.}
\centering
\includegraphics[scale=0.41]{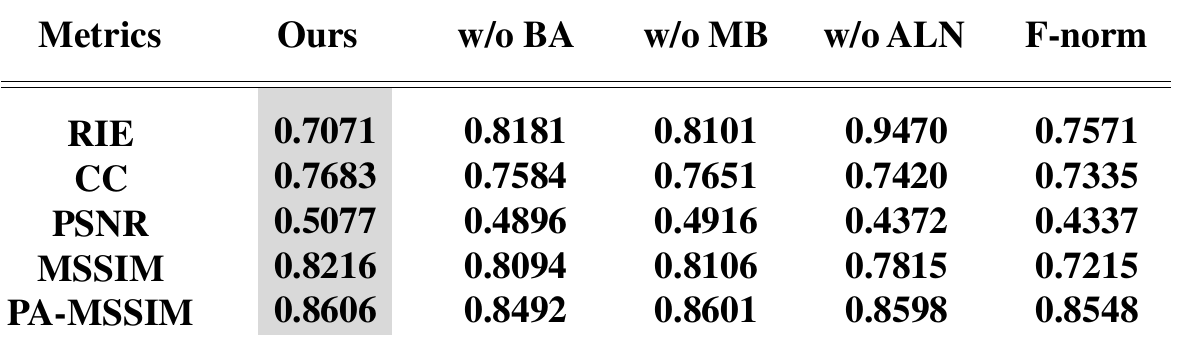}
\label{table_ablation}
\end{table}

\subsection{Simulation Results}

\subsubsection{Performance Comparison} Fig.~\ref{simulation_results} shows the mfEIT image reconstruction results for three simulation cases using the proposed MAIP algorithm and five other image reconstruction algorithms, based on noise-free simulation data.. For Case 1, it is evident that the model-based supervised learning method based on the MMV model (MMV-Net) incorrectly reconstructs the inter-frequency differences. This is because such methods rely on data-driven training and often struggle to generalize effectively, inevitably learning the inter-frequency correlations embedded in the training dataset. Therefore, when the inter-frequency correlations in the test data deviate from those in the training data, the model produces reconstructions with incorrect inter-frequency correlations. In Case 2, the MAIP algorithm significantly outperforms the other five algorithms in reconstructing multiple complex targets. It demonstrates superior structure preservation, more accurate inter-frequency correlations, and produces fewer artefacts. In contrast, SMV-based methods underperform mainly in reconstructing lower conductivity contrasts, especially at ${f_{4}}$. Similar results can also be observed in Case 3. Overall, MMV-based methods exhibit better structural consistency in multi-frequency reconstructed images compared to SMV-based methods.

Table~\ref{metrics_1} provides the average quantitative metrics for the six algorithms across all cases. The proposed MAIP algorithm consistently outperforms the other methods on all metrics. Specifically, MAIP achieves lower RIE values, indicating higher pixel-level accuracy. Moreover, the higher PSNR achieved by our method indicates better noise resistance, which suggests clearer reconstructions with fewer artefacts. Furthermore, the higher CC and MSSIM values indicate that MAIP provides more accurate structural reconstructions. Lastly, MAIP records the highest PA-MSSIM, highlighting its superior structural consistency across different observation frequencies. Fig.\ref{GtvsPred}a shows the selected 4x4 region of Case 2 for further analysis. In Fig.\ref{GtvsPred}b, the average conductivity curves for the ground truth and reconstructed images are compared across frequencies. The MAIP algorithm (labeled "Ours") closely follows the ground truth, particularly at lower frequencies.%Fig.~\ref{GtvsPred}(b) (bottom) presents Pearson correlation, cosine similarity, and MSE between the reconstructions and the ground truth. While the other two MMV-based methods, ADMM-MMV and MMV-Net, also show good correlation with the ground truth compared to SMV-based methods, ADMM-MMV exhibits a significantly higher MSE. MMV-Net performs well overall, with relatively low MSE and good similarity metrics, though it still falls short of MAIP.

Overall, the simulation results indicate that MAIP outperforms the other five algorithms across all three cases. Its advantages include accurate inter-frequency correlations, more precise shape reconstruction, and fewer artefacts. However, compared to model-based supervised learning methods, both traditional model-based iterative methods and model-based unsupervised learning methods tend to produce slight edge noise in reconstructed images, particularly for imaging objects with sharp edges, such as triangles or rectangles. This issue arises due to the ill-posed nature of EIT inverse problem. In contrast, supervised learning methods are able to effectively eliminate such boundary noise by leveraging data-driven training. Additionally, accurately reconstructing the conductivity differences between multiple objects with similar conductivity remains a significant challenge.

\subsubsection{Convergence and Noise Resistance} 
% \begin{figure*}[!t]	\centerline{\includegraphics[scale=0.36]{figures/noise_resistance2.png}}
% 	\caption{Simulation results from Case 2: (a) Training loss curves of our algorithm under different noise levels; (b) Effect of noise on quantitative metrics; (c) The bloxplot of quantitative metrics.}
% 	\label{noise_resistance}
% \end{figure*}

To verify the convergence and noise resistance of our method, we introduced varying levels of Gaussian noise into the voltage data, generating noisy voltage measurements with SNRs ranging from 10 dB to 90 dB, and conducted our simulations based on these datasets. Fig.~\ref{loss_noise_curve1} displays the convergence curves of the training loss at different noise levels. The results confirm that our method has smooth convergence across various noise levels. Fig.~\ref{noise_resistance}a presents the changes in five quantitative metrics (RIE, CC, PSNR, MSSIM and PA-MSSIM) as the SNR varies, while Fig.~\ref{noise_resistance}b gives a boxplot of these metrics. The results indicate that our method exhibits excellent noise resistance, achieving stable reconstruction results for SNRs above 20 dB.

\subsubsection{Ablation study}
Fig.~\ref{ablation} and Table~\ref{table_ablation} give the results of the ablation study conducted on Case 2. We compared the mfEIT reconstruction results after removing the Branch Attention (BA) module, the Multi-Branch (MB) structure, and Adaptive Layer Normalization (ALN) (replaced with batch normalization). Additionally, we evaluated the impact of substituting the loss function $\ell_1$ norm with the Frobenius norm (F-norm).

The ablation study results indicate that removing the BA module or using a single-branch structure causes a significant increase in RIE and a slight decrease in all other metrics, suggesting that both the BA module and the multi-branch structure effectively enhance the quality of mfEIT reconstruction. Additionally, reconstructions without the BA module exhibit the lowest PA-MSSIM, suggesting that BA also enhances the algorithm’s ability to improve inter-frequency structural consistency across different frequencies. Furthermore, we found that the introduction of layer normalization significantly improves the image quality, particularly in accurately capturing the inter-frequency conductivity differences. Lastly, using $\ell_1$ loss as the loss function in the iterative optimization process significantly enhances the ability to reconstruct accurate shapes compared to Frobenius loss.

\begin{figure}[!t]
	\centerline{\includegraphics[scale=0.25]{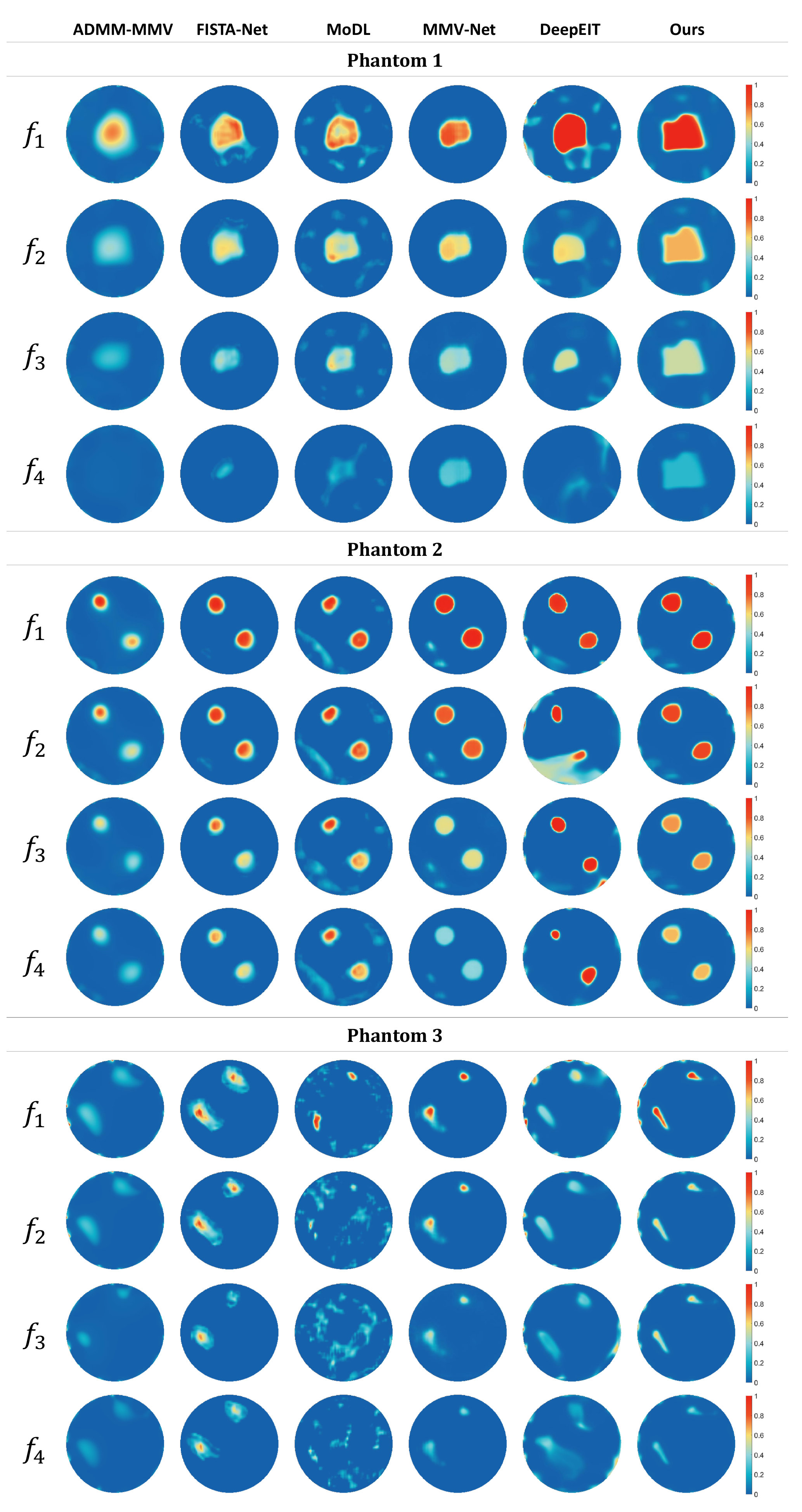}}
	\caption{Comparison of the MAIP algorithm with five SOTA image reconstruction algorithms using real-world data. }
	\label{experimental_results}
\end{figure}

\subsection{Experimental Results}
Fig.~\ref{experimental_results} compares the performance of our MAIP algorithm with five SOTA image reconstruction algorithms on three sets of experimental data, while Table \ref{table_realworld} provides the average quantitative metrics (MSSIM and PA-MSSIM) for these real-world experiments.  The three Phantoms are detailed in Fig.~\ref{real-world_exp}. Our MAIP algorithm achieved the highest MSSIM scores across all channels, notably excelling in ${f_{4}}$ with an MSSIM of 0.8275, and consistently provided the best average MSSIM and PA-MSSIM values of 0.8172 and 0.8572, respectively, indicating superior performance compared to the other methods.

Both FISTA-Net and MMV-Net also performed well in reconstructing targets with lower conductivity contrasts, such as those at ${f_{3}}$ and ${f_{4}}$. In contrast, MoDL failed to reconstruct the shapes in Phantom 3, with the failure beginning as early as ${f_{2}}$. Moreover, only MMV-Net and the proposed MAIP achieved good inter-frequency structural consistency and were able to consistently provide a clear trend of conductivity values relative to frequency. Notably, MMV-Net, influenced by the inter-frequency variations in the training data, exhibited identical inter-frequency correlations across all three phantoms. Our MAIP avoided this issue. Additionally, MAIP produced more accurate shapes and fewer artefacts compared to the other methods. This was particularly evident in imaging two zebrafish in Phantom 3, which was the most challenging of the three experiments. Only MAIP successfully reconstructed the relatively accurate shapes of the zebrafish across all frequencies. The Phantom 2 results highlight the challenge of distinguishing between the conductivities of different objects in multi-target imaging, indicating an area for further improvement.

Overall, the MAIP algorithm consistently outperforms the five other image reconstruction algorithms across all three experimental cases. It shows notable strengths in reconstructing targets with lower conductivity contrasts, maintaining inter-frequency structural consistency, and accurately capturing complex shapes, such as the zebrafish in Phantom 3. Compared to other methods, MAIP demonstrates an ability to adapt to varying inter-frequency correlations and reduce artefacts, making it a promising option for the challenging mfEIT reconstruction task. 

\begin{table}[!t]
\caption{Quantitative Comparisons ( MSSIM and PA-MSSIM) for Real-world Experiment Results.}
\centering
\includegraphics[scale=0.35]{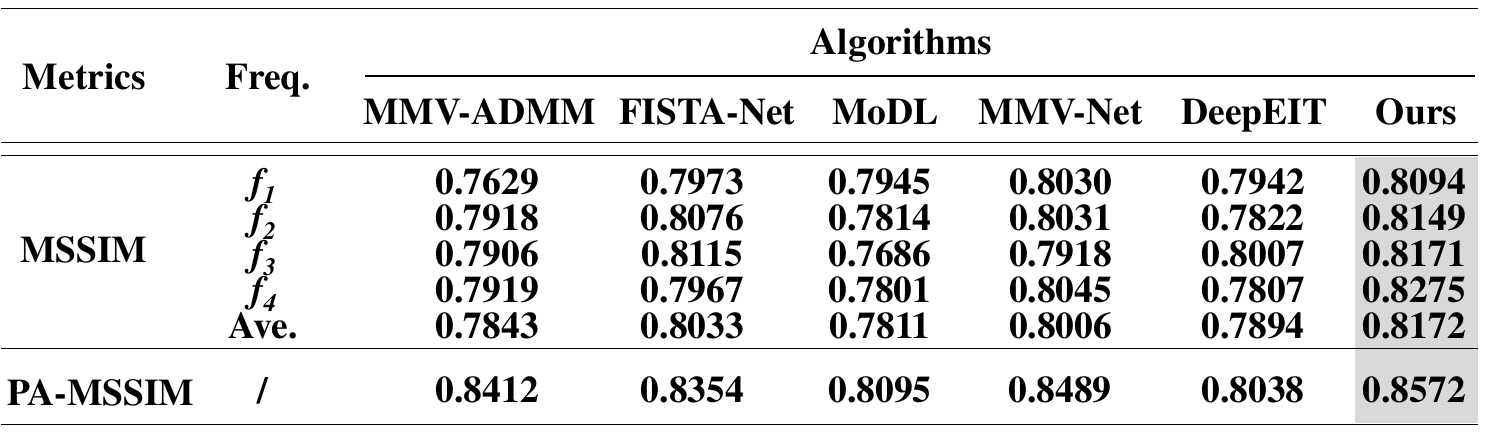}
\label{table_realworld}
\end{table}

\section{CONCLUSION}
We presented a model-based unsupervised learning method for mfEIT image reconstruction, named Multi-Branch Attention Image Prior (MAIP). This approach leverages a Multi-Branch Attention Network (MBA-Net) to represent the multi-frequency conductivity distributions. Notably, our method requires no training data for optimizing the neural network parameters. The deep architecture of MBA-Net captures complex intra-frequency correlations, while the multi-branch structure and the branch attention mechanism assist in accurately reconstructing inter-frequency correlations. Our method demonstrates smooth convergence and strong noise resistance, with performance comparable to state-of-the-art supervised learning algorithms, as evidenced by simulations and real-world experiments. Future work includes extending our approach to 3D mfEIT imaging and other tomographic imaging tasks.

% \clearpage
\FloatBarrier
\bibliographystyle{IEEEtran}
\bibliography{reference}

\end{document}